\documentclass[12pt]{iopart}
\usepackage{graphicx}
\usepackage{iopams}
\usepackage{amssymb}
\usepackage{multirow}
\usepackage{xcolor}

\begin{document}
\title{Constrains on $s$ and $d$ components of electron boson coupling constants
in one band d wave Eliashberg theory for high Tc superconductors}
\author{G.A.~Ummarino}
\ead{giovanni.ummarino@polito.it}
\address{Istituto di Ingegneria e Fisica dei Materiali,
Dipartimento di Scienza Applicata e Tecnologia, Politecnico di
Torino, Corso Duca degli Abruzzi 24, 10129 Torino, Italy; National Research Nuclear University MEPhI (Moscow Engineering Physics Institute),
Kashira Hwy 31, Moskva 115409, Russia}
\begin{abstract}
The phenomenology of overdoped high $T_{c}$ superconductors can be described by a one band $d$ wave Eliashberg theory where the mechanism of superconducting coupling is mediated by antiferromagnetic spin fluctuations and whose characteristic energy $\Omega_{0}$ scales with $T_{c}$ according to the empirical law $\Omega_{0} = 5.8 k_{B}T_{c}$. This model presents universal characteristics that are independent of the critical temperature such as the link between the $s$ and $d$ components of electron boson coupling constants and the invariance of the ratio $2\Delta/k_{B}T_{c}$. This situation arises from the particular structure of Eliashberg's equations which, despite being non-linear equations, present solutions with these simple properties.
\end{abstract}
%
%
\maketitle
\section{INTRODUCTION}
Eliashberg's theory \cite{Eliashberg} was born as a generalization of the BCS theory to explain some anomalies in the experimental data concerning lead. Subsequently it was seen that the theory can be successfully applied to explain the experimental data of practically almost all superconducting materials \cite{ummarinorev,marsigliorev}: first of all low $T_{c}$ phononic superconductors \cite{carbirev}, then magnesium diboride \cite{magdib1,magdib2}, graphite intercalated compound $CaC_{6}$ \cite{graphite}, iron-based superconductors \cite{iron1,iron2,iron3,iron4,iron5}. This theory can be applied to describe particular systems such as proximized systems \cite{prox1} and field effect junctions \cite{proxE1,proxE2,proxE3}.
For what concerns the high $T_{c}$ superconductors \cite{HTCS2001EPJ,new1,new2,new3,new5}, their properties strongly depend on their oxygen content. It is possible to identify three different regimes: under, optimal and overdoping. While the discussion is still open as regards the underdoping regime, it is almost certain that the fundamental mechanism in the optimal and over regime is due to antiferromagnetic spin fluctuactions and especially in the over regime, the experimental data can be described satisfactorily by one band $d$-wave Eliashbeg's theory  \cite{HTCSumma,Dwave1}. Detailed studies are present in literature on cuprates and precisely on tunneling spectra that can be reproduced by using in the framework of $d$-wave Eliashbeg's theory \cite{ema1,ema2,ema3}.
In this paper we provide an extensive investigation of the consequences of a different symmetry of coupling in the two components of self energy: the renormalisation function $Z(i\omega_{n})$ (s-wave symmetry) and the gap function $\Delta(i\omega_{n})$ (d-wave symmetry) and if some link exists between them. We focus here on physical quantities which can be evaluated in the imaginary axis formalism.
Furthermore, it has been experimentally determined that, in cuprates, a link \cite{Yu2009Nature} exists between magnetic resonance energy $\Omega_{0}$ and critical temperature. So we will study the properties of one band $d$-wave Eliashbeg's theory where a fundamental role will be played by the assumption that the representative energy $\Omega_{0}$ of these systems is related to the critical temperature by a universal relationship \cite{Yu2009Nature} $\Omega_{0} = 5.8k_{B}T_{c}$. This assumption represents a very strong constraint in correlating the values of the two electron boson coupling constants $\lambda_{d}$ and $\lambda_{s}$. For each value of $\lambda_{s}$ we will look for the value of $\lambda_{d}$ which exactly reproduces the $T_{c}$ of the superconductor and we will study which relation exists between the $d$ and $s$ components of electron boson coupling constant.
Finally we will see that this model has the particular property that the relationship between the gap and the critical temperature ($\frac{2\Delta_{d}}{k_{B}T_{c}}$) is independent of the particular value of the critical temperature.
\label{intro}
%
\section{MODEL}
\label{sec:model}
The one band d-wave Eliashberg equations \cite{Dwave1,Dwave2,Dwave3,Dwave4,Dwave5,Dwave6,Dwave7} are two coupled equations: one for the gap $\Delta(i\omega_{n},\phi)$ and one for the renormalization functions $Z(i\omega_{n},\phi)$. These equations, in the imaginary axis representation (here $\omega_{n}$ denotes the Matsubara frequencies), when the Migdal theorem holds \cite{Migdal}, are:

\begin{eqnarray}
\omega_{n}Z(\omega_{n},\phi)=\omega_{n}+\pi T
\sum_{m}\int_{0}^{2\pi}\frac{d\phi'}{2\pi}\Lambda(\omega_{n},\omega_{m},\phi,\phi')N_{Z}(\omega_{m},\phi')
\end{eqnarray}
\begin{eqnarray}
&&Z(\omega_{n},\phi)\Delta(\omega_{n},\phi)=\pi T
\sum_{m}\int_{0}^{2\pi}\frac{d\phi'}{2\pi}[\Lambda(\omega_{n},\omega_{m},\phi,\phi')-\mu^{*}(\phi,\phi')
\big]\times\nonumber\\
&&\times\Theta(\omega_{c}-|\omega_{m}|)N_{\Delta}(\omega_{m},\phi')
\label{eq:EE2}
\end{eqnarray}
where $\Theta(\omega_{c}-\omega_{m})$ is the Heaviside function, $\omega_{c}$ is a
cut-off energy and
\begin{eqnarray}
\Lambda(\omega_{n},\omega_{m},\phi,\phi')=2\int_{0}^{+\infty}\Omega d\Omega
\alpha^{2}F(\Omega,\phi,\phi')/[(\omega_{n}-\omega_{m})^{2}+\Omega^{2}]
\end{eqnarray}
\begin{eqnarray}
N_{Z}(\omega_{m},\phi)=
\frac{\omega_{m}}{\sqrt{\omega^{2}_{m}+\Delta(\omega_{m},\phi)^{2}}}
\end{eqnarray}
\begin{eqnarray}
N_{\Delta}(\omega_{m},\phi)=
\frac{\Delta(\omega_{m},\phi)}{\sqrt{\omega^{2}_{m}+\Delta(\omega_{m},\phi)^{2}}}
\end{eqnarray}
We assume \cite{ummarinorev,Dwave1,Dwave2,Dwave3,Dwave4,Dwave5,Dwave6,Dwave7} that the electron boson spectral
function $\alpha^{2}(\Omega)F(\Omega,\phi,\phi')$ and the Coulomb
pseudopotential $\mu^{*}(\phi,\phi')$ at the lowest order to contain
separated $s$ and $d$-wave contributions,
\begin{equation}
\alpha^{2}F(\Omega,\phi,\phi')=\lambda_{s}\alpha^{2}F_{s}(\Omega)
+\lambda_{d}\alpha^{2}F_{d}(\Omega)\sqrt{2}cos(2\phi)\sqrt{2}cos(2\phi')
\end{equation}
\begin{equation}
\mu^{*}(\phi,\phi')=\mu^{*}_{s}
+\mu^{*}_{d}\sqrt{2}cos(2\phi)\sqrt{2}cos(2\phi')
\end{equation}
as well as the self energy functions:
\begin{equation}
Z(\omega_{n},\phi)=Z_{s}(\omega_{n})+Z_{d}(\omega_{n})cos(2\phi)
\end{equation}
\begin{equation}
\Delta(\omega_{n},\phi)=\Delta_{s}(\omega_{n})+\Delta_{d}(\omega_{n})cos(2\phi)
\end{equation}
We put the factor $\sqrt{2}$ inside the definition of $\Delta_{d}(\omega_{n})$ because, experimentally, the peak of the density of state is, usually, identified with $\Delta_{d}(\omega_{n=0})$ while, as we will see, $Z_{d}(\omega_{n})$ is always zero.
The spectral functions $\alpha^{2}F_{s,d}(\Omega)$ are normalized in the way that $2\int_{0}^{+\infty}\frac{\alpha^{2}F_{s,d}(\Omega)}{\Omega}d\Omega=1$
and, of course, in this model the renormalization function is pure $s$-wave ($Z(\omega_{n},\phi)=Z_{s}(\omega_{n})$) while
the gap function is pure $d$-wave ($\Delta(\omega_{n},\phi)=\Delta_{d}(\omega_{n})cos(2\phi)$).
We consider just solutions of the Eliashberg equations in pure $d$-wave form because this is the indication of the experimental data. This means that the $s$ component of the gap function is zero and this situation happens because, usually, \cite{varelo}, $\mu^{*}_{s}>>\mu^{*}_{d}$). In the more general case, in principle, the gap function has $d$ and $s$ components. The renormalization function $Z(\omega,\phi')=Z_{s}(\omega)$ has just the $s$ component because
the equation for $Z_{d}(\omega_{n})$ is a homogeneous integral equation with just the solution
$Z_{d}(\omega_{n})=0$ \cite{zetad}. For simplicity we also assume that $\alpha^{2}F_{s}(\Omega)=\alpha^{2}F_{d}(\Omega)$ and that the
spectral functions is the difference of two Lorentzian, i.e. $\alpha^{2}F_{s,d}(\Omega)=C[L(\Omega+\Omega_{0},\Upsilon)-L(\Omega-\Omega_{0},\Upsilon)]$
where $L(\Omega\pm\Omega_{0},\Upsilon))=[(\Omega\pm\Omega_{0})^{2}+(\Upsilon)^{2}]^{-1}$,
$C$ is the normalization constant necessary to obtain $2\int_{0}^{\infty}\frac{\alpha^{2}F_{s,d}(\Omega)}{\Omega}d\Omega=1$, $\Omega_{0}$ and $\Upsilon$ are the peak
energy and half-width, respectively. The half-width is
$\Upsilon=\Omega_{0}/2$. This choice of the shape of spectral function and the fact that $\alpha^{2}F_{s}(\Omega)=\alpha^{2}F_{d}(\Omega)$, is a good approximation of the true spectral function \cite{a2fd}
connected with antiferromagnetic spin fluctuations. The same thing also happens in the case of iron pnictides \cite{ummarinoiron}.
In any case, even making different choices for $\Upsilon $ the link between $\lambda_{d}$ and $\lambda_{s}$) remains the same but change (very little) the coefficients of the linear fit.
The cut-off energy is $\omega_{c}=1000$ meV and the maximum quasiparticle energy is $\omega_{max}=1100$ meV.
In first approximation we put $\mu^{*}_{d}=0$ (if the $s$ component of the gap is zero the value of $\mu^{*}_{s}$ is irrelevant).
Now we fix the critical temperature and for any value of $\lambda_{s}$ we seek the value of $\lambda_{d}$ that exactly reproduce the initial fixed critical temperature.
After, via Pad\`{e} approximants \cite{Vidberg}, we calculate the low-temperature value ($T=T_{c}/10$ K) of the gap because, in presence of a strong coupling
interaction, the value of $\Delta_{d}(i\omega_{n=0})$ obtained by solving the imaginary-axis
Eliashberg equations can be very different from the value of $\Delta_{d}$ obtained from the real-axis Eliashberg equations \cite{Dwave3}.
\section{RESULTS AND DISCUSSION}
We fix three different critical temperatures ($70$ K, $90$ K and $110$ K) and, for any particular critical temperature, we choose different values of $\lambda_{s}$ and determine which value of $\lambda_{d}$ exactly reproduces the chosen critical temperature by numerical solution of Eliashberg equations. In figure 1 we can see that the three curves $\lambda_{d}$ versus $\lambda_{s}$ are coincident. In the inset of figure 1 it is shown the linear fit of these results. We obtain a linear link between $\lambda_{d}$ and $\lambda_{s}$
\begin{equation}
\lambda_{d}=0.616\lambda_{s}+0.732
\end{equation}
These results are general and dont depend from the particular shape of the electron-boson spectral function.
If we change the shape of the electron-boson spectral function and we choose, for example, $\alpha^{2}F_{s,d}(\Omega)=0.5\Omega_{0}\delta(\Omega-\Omega_{0})$ we find that the linear link between $\lambda_{d}$ and $\lambda_{s}$ changes very little and becomes $\lambda_{d}=0.575\lambda_{s}+0.655$.
Even the introduction of a Coulomb potential different from zero, as we have verified, does not involve a substantial modification of our results.
In principle it is possible to obtain this result (linear link between $\lambda_{s}$ and $\lambda_{d}$) in a more simple but less general way. In fact a similar conclusion relative to linear connection between $\lambda_{s}$ and $\lambda_{d}$ may also be derived from the analysis of the approximate MacMillan formula for $T_{c}$ \cite{MacMillan} generalized to $d$-wave case \cite{carbid}:
\begin{equation}
k_{B}T_{c}=\Omega_{0}exp(-\frac{1+\lambda_{s}}{2\lambda_{d}})
\end{equation}
The problem is that the MacMillan equation works just in weak coupling regime.
Now we solve, for each couple of $\lambda_{d}$ and $\lambda_{s}$ values, the Eliashberg equations at $T=T_{c}/10$ and after, via Pade we calculate the value of superconductive gap (the energy of the density of states peak). In figure 2 the rates $\frac{2\Delta_{d}}{k_{B}T_{c}}$ are shown for three systems with different critical temperatures ($70$ K, $90$ K and $110$ K). The curves are exactly coincident.
We have also studied what happens when the ratio $\frac{\Omega_{0}}{k_{B}T_{c}}$ is equal to two as in the case of the heavy fermion \cite{UPdAl} $UPd_{2}Al_{3}$ with $T_{c}=2$ K that it could represent an extreme situation. In this case the link remains linear and becomes $\lambda_{d}=0.880\lambda_{s}+0.966$ as it is possible see in the inset of figure 2.
Finally in the case of extreme strong coupling ($\frac{\Omega_{0}}{k_{B}T_{c}}<<1$) it is possible to demonstrate in an analytical way, following the calculus of ref 26, when $\frac{\lambda_{s}}{2\lambda_{d}}>1$, that $\lambda_{d}\approx \lambda_{s}$ i.e. the link remains linear.
\label{sec:results}
\section{CONCLUSIONS}
In this article it has been shown that one band $d$-wave Eliashbeg's theory presents universal aspects as the linear link between $\lambda_{d}$ and $\lambda_{s}$ or the values of $2\Delta_{d}/k_{B}T_{c}$ that are independent of the particular critical temperature.
These universal aspects are relate to the assumption that the typical bosonic energy is correlated to the critical temperature as shown by experimental data ($\Omega_{0}=5.8k_{B}T_{c}$).
We here proved that in a fully numerical solution of the Eliashberg equation such linear link hold with great accuracy.
A generalization and development of our results can be obtained by explicitly considering the momentum dependence of the self-energy without average on the Fermi surface as was done by Kamila A. Szewczyk et al \cite{kamila}. Obviously we would include in the calculations, unlike them, as we have done now, the link, observed experimentally, between the critical temperature and the representative energy of the bosonic spectrum.
\label{sec:conclusions}
\ackn
The author acknowledges support from the MEPhI Academic Excellence Project (Contract No. 02.a03.21.0005).\\

\newpage
\begin{figure}
\begin{center}
\includegraphics[keepaspectratio, width=\columnwidth]{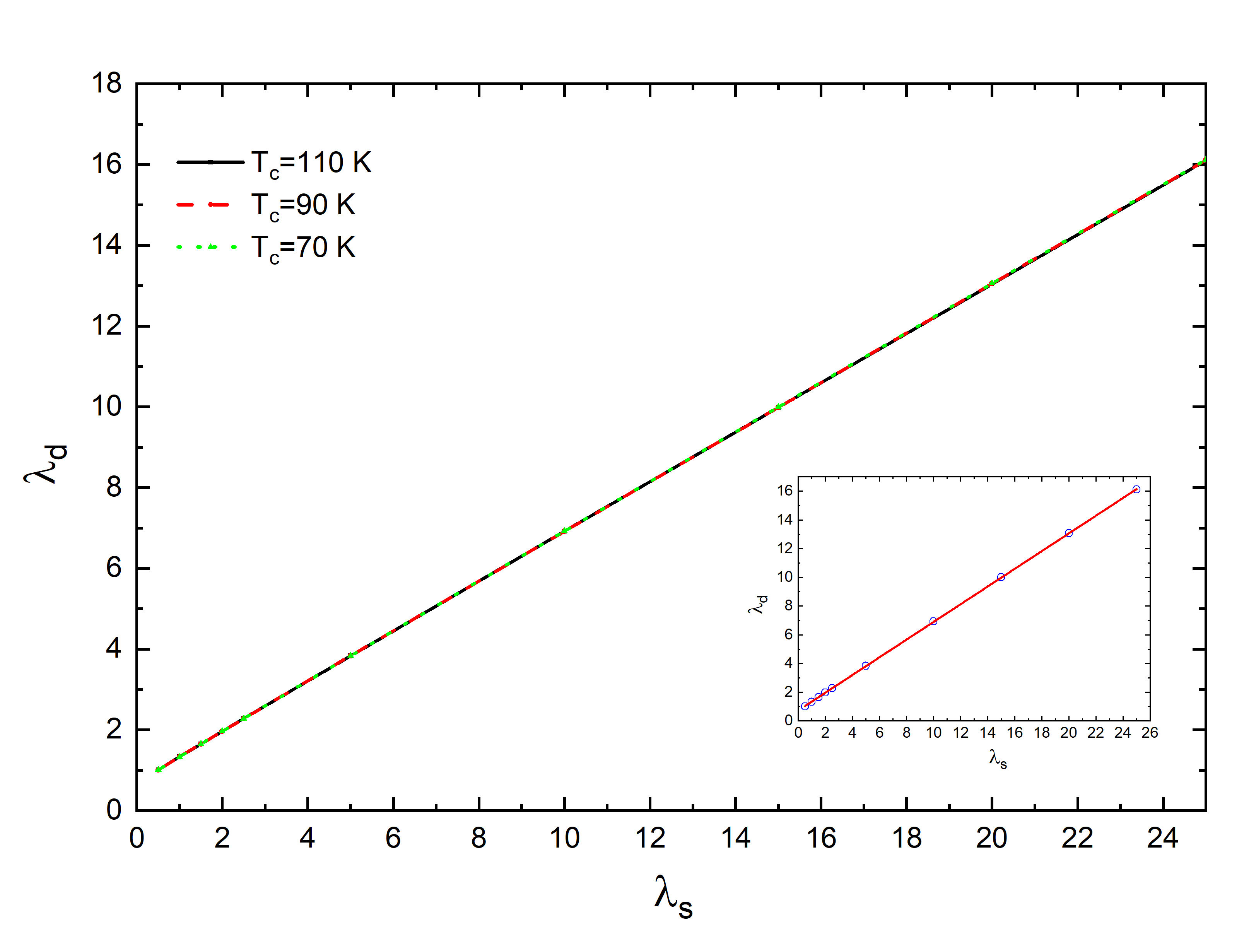}
\vspace{-5mm} \caption{(Color online)
 $\lambda_{d}$ versus $\lambda_{s}$ for three different critical temperatures: $T_{c}=70$ K (green point line), $T_{c}=90$ K (red dash line) and $T_{c}=110$ K (black solid line). In the inset the linear fit (solid line) of the $T_{c}=70$ K (open dark blue circles) case is shown.
 }\label{Figure1}
\end{center}
\end{figure}
\newpage
\begin{figure}
\begin{center}
\includegraphics[keepaspectratio, width=\columnwidth]{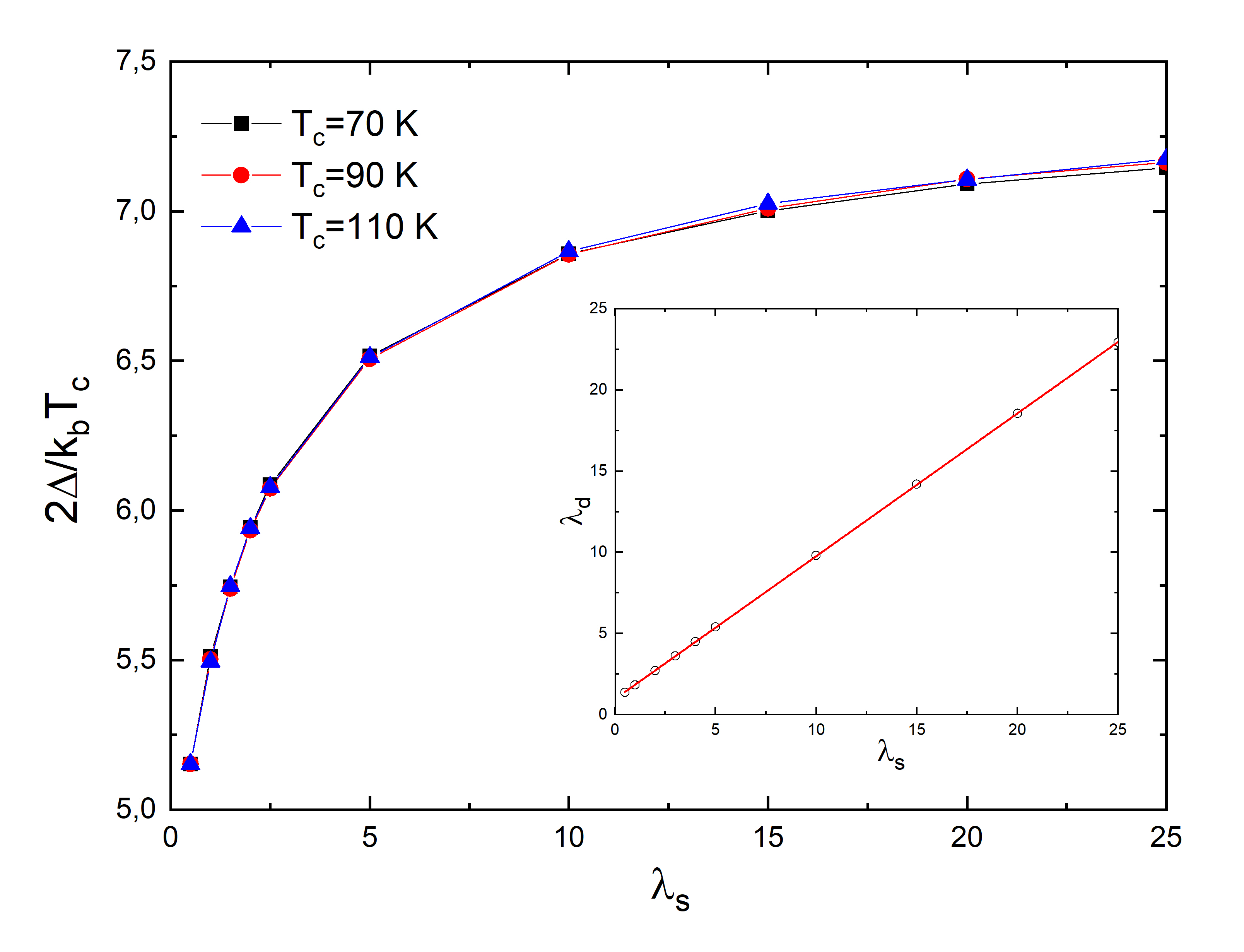}
\vspace{-5mm} \caption{(Color online)
 $|\Delta_{i}|/k_{B}T_{c}$ for $T_{c}=70$ K (green up filled triangles circles), $T_{c}=90$ K (red filled circles) and $T_{c}=110$ K (black filled squares) versus $|\lambda_{s}|$. The lines are guides for eyes.
 In the inset $|\lambda_{d}|$ versus $|\lambda_{s}|$ when $T_{c}=2$ K and $\Omega_{0}=2k_{B}T_{c}$ is shown (black open circles) with the linear fit (solid red line).
 }\label{Figure2}
\end{center}
\end{figure}

\begin{thebibliography}{99}
\small{

\bibitem{Eliashberg} G.M. Eliashberg, \textit{Sov. Phys. JETP} \textbf{3}, 696 (1963).
%
\bibitem{ummarinorev} G.A. Ummarino, Eliashberg Theory. In: \textit{Emergent Phenomena in Correlated Matter}, edited by E. Pavarini, E. Koch, and U. Schollw\"{o}ck, Forschungszentrum J\"{u}lich GmbH and Institute for Advanced Simulations, pp.13.1-13.36 (2013) ISBN 978-3-89336-884-6.
    %
\bibitem{marsigliorev} F. Marsiglio, \textit{Annals of Physics} \textbf{417}, 168102 (2020).
%
\bibitem{carbirev} J.P. Carbotte, \textit{Rev. Mod. Phys.} \textbf{62}, 1027 (1990).
%
\bibitem{magdib1} D. Daghero, R.S. Gonnelli, G.A. Ummarino, S.M. Kazakov, J. Karpinski, V.A. Stepanov, J. Jun,
\textit{Physica C} \textbf{385}, 255 (2003).
%
\bibitem{magdib2} D. Daghero, A. Calzolari, G.A. Ummarino, M. Tortello, and R. S. Gonnelli, V.A. Stepanov, C. Tarantini, P. Manfrinetti, E. Lehmann, \textit{Phys. Rev. B} \textbf{74}, 174519 (2006).
%
\bibitem{graphite} A. Sanna, S. Pittalis, J.K. Dewhurst, M. Monni, S. Sharma, G. Ummarino, S. Massidda, and E.K.U. Gross, \textit{Phys. Rev. B} \textbf{85}, 184514 (2012).
%
\bibitem{iron1} D. Torsello, G.A. Ummarino, L. Gozzelino, T. Tamegai, G. Ghigo, \textit{Phys. Rev. B} \textbf{99}, 134518 (2019).
%
\bibitem{iron2} D. Torsello, G.A. Ummarino, J. Bekaert, L. Gozzelino, R. Gerbaldo,
M.A. Tanatar, P.C. Canfield , R. Prozorov, and G. Ghigo, \textit{Phys. Rev. Appl.} \textbf{13}, 064046 (2020).
%
\bibitem{iron3} G. Ghigo, G.A. Ummarino, L. Gozzelino, and T. Tamegai, \textit{Phys. Rev. B} \textbf{96}, 014501 (2017).
%
\bibitem{iron4} D. Torsello, K. Cho, K.R. Joshi, S. Ghimire, G.A. Ummarino, N. M. Nusran, M.A. Tanatar, W.R. Meier, M. Xu, S.L. Budko,
P.C. Canfield, G. Ghigo, and R. Prozorov, \textit{Phys. Rev. B} \textbf{100}, 094513 (2019).
%
\bibitem{iron5} G.A. Ummarino, \textit{Magnetochemistry} \textbf{9}, 28 (2023).
%
\bibitem{prox1} G.A. Ummarino, \textit{Physica C} \textbf{568}, 1353566 (2020).
%
\bibitem{proxE1} G.A. Ummarino, E. Piatti, D. Daghero, R.S. Gonnelli, Irina Yu. Sklyadneva, E.V. Chulkov and R. Heid, \textit{Physical Review B} \textbf{96}, 064509 (2017).
%
\bibitem{proxE2} G.A. Ummarino and D. Romanin, \textit{Phys. Status Solidi B} \textbf{2020}, 1900651 (2020).
%
\bibitem{proxE3} G.A. Ummarino and D. Romanin, \textit{J. Phys.: Condens. Matter} \textbf{31}, 024001 (2019).
%
\bibitem{HTCS2001EPJ} R.S. Gonnelli, A. Calzolari, D. Daghero, L. Natale, G.A. Ummarino, V.A. Stepanov, M. Ferretti, \textit{European Physical Journal B} \textbf{22}, 41 (2001).
    %
\bibitem{new1} Sima Alikhanzadeh-Arani, Masoud Salavati-Niasari, Mohammad Almasi-Kashi, \textit{Physica C} \textbf{488}, 30 (2013).
    %
    \bibitem{new2} Sima Alikhanzadeh-Arani, Masoud Salavati-Niasari, Mohammad Almasi-Kashi, \textit{Journal of Inorganic and Organometallic Polymers and Materials} \textbf{22}, 1081 (2012).
    %
\bibitem{new3} Sima Alikhanzadeh-Arani, Mahboubeh Kargar, Masoud Salavati-Niasari, \textit{Journal of Alloys and Compounds} \textbf{614}, 35 (2014).
 %
 %
\bibitem{new5} Mahboubeh Kargar, Sima Alikhanzadeh-Arani,
Zahra Pezeshki-Nejad, Masoud Salavati-Niasari, \textit{Journal of Superconductivity and Novel Magnetism} \textbf{28}, 13 (2015).
%
\bibitem{HTCSumma} G.A. Ummarino, \textit{Condens. Matter} \textbf{6}, 13 (2021).
   %
\bibitem{Dwave1} C.T. Rieck, D. Fay, L. Tewordt, \textit{Phys. Rev. B} \textbf{41}, 7289 (1989).
%
\bibitem{ema1} C. Jiang, J. P. Carbotte, and R. C. Dynes, \textit{Phys. Rev. B} \textbf{47}, 5325 (1993).
%
\bibitem{ema2} J.F. Zasadzinski, L. Coffey, P. Romano, and Z. Yusof, \textit{Phys. Rev. B} \textbf{68}, 180504(R) (2003).
%
\bibitem{ema3} O. Ahmadi, L. Coffey, J.F. Zasadzinski, N. Miyakawa, and L. Ozyuzer
\textit{Phys. Rev. Lett} \textbf{106}, 167005 (2011).
%
\bibitem{Yu2009Nature} G. Yu, Y. Li, E. M. Motoyama and M. Greven, \textit{Nature Physics} \textbf{5}, 873 (2009).
%
%
\bibitem{Ghigo2017scirep} G. Ghigo, G.A. Ummarino, L. Gozzelino, R. Gerbaldo, F. Laviano, D. Torsello, T. Tamegai,
\textit{Sci. Rep.} \textbf{7}, 13029 (2017).
 %
 \bibitem{Migdal} G.A.Ummarino and R.S. Gonnelli, \textit{Phys. Rev. B} \textbf{56}, 14279 (1997).
%
\bibitem{Dwave2}  G.A.Ummarino and R.S. Gonnelli, \textit{Physica C} \textbf{328}, 189 (1999).
 %
\bibitem{Dwave3}  G.A.Ummarino and R.S. Gonnelli, \textit{Physica C} \textbf{341-348}, 295 (2000).
%
\bibitem{Dwave4} G.A. Ummarino, D. Daghero and R.S. Gonnelli, \textit{Physica C} \textbf{377}, 292 (2002).
%
\bibitem{Dwave5} E. Cappelluti, G.A. Ummarino, \textit{Phys. Rev. B} \textbf{76}, 104522 (2007).
%
\bibitem{Dwave6} F. Jutier, G.A. Ummarino, J.C. Griveau, F. Wastin, E. Colineau, J. Rebizant, N. Magnani, and R. Caciuffo, \textit{Phys. Rev. B} \textbf{77}, 024521 (2008).
    %
\bibitem{Dwave7} G.A. Ummarino, R. Caciuffo, H. Chudo and S. Kambe, \textit{Phys. Rev. B} \textbf{82}, 104510 (2010).
%
\bibitem{varelo} Georgios Varelogiannis, \textit{Solid State Communications} \textbf{107}, 427 (1998).
%
\bibitem{a2fd} Jin Mo Bok, Jong Ju Bae, Han-Yong Choi, Chandra M. Varma, Wentao Zhang,
Junfeng He, Yuxiao Zhang, Li Yu, X. J. Zhou, \textit{Sci. Adv.} \textbf{2}, 1501329 (2016)
%
\bibitem{ummarinoiron} G.A. Ummarino, \textit{Phys. Rev. B} \textbf{83}, 092508 (2011).
%
\bibitem{zetad} K. A. Musaelian, J. Betouras, A. V. Chubukov, and R. Joynt, \textit{Phys. Rev. B} \textbf{53}, 3598 (1996).
%
\bibitem{Vidberg} H. Vidberg and J. Serene, \textit{J. Low Temp. Phys.} \textbf{29} 29, 179 (1977).
%
\bibitem{MacMillan} W.L. McMillan, \textit{Phys. Rev.} \textbf{167}, 331 (1968).
%
\bibitem{carbid} J. Hwang, E. Schachinger, J.P. Carbotte, F. Gao, D.B. Tanner, and T. Timusk, \textit{Phys. Rev. Lett.} \textbf{100}, 137005 (2008).
    %
\bibitem{UPdAl} N.K. Sato, N. Aso, K. Miyake, R. Shiina, P. Thalmeier, G. Varelogiannis, C. Geibel, F. Steglich, P. Fulde, T. Komatsubara,
\textit{Nature} \textbf{410}, 340 (2001).
%
\bibitem{kamila} Kamila A. Szewczyk, Radosław Szczesniak, and Dominik Szczesniak, \textit{Annalen der Physik} \textbf{530}, 1800139 (2018).
%

}\end{thebibliography}
\end{document}